\documentclass[twocolumn,amsmath,amssymb,10pt]{revtex4}

\usepackage{color}
\usepackage{framed}

\usepackage{graphicx}

\usepackage{amsmath}

\begin{document}            

\title{Random Networks with Tunable Degree Distribution and Clustering}

\author{Erik Volz}
\affiliation{Cornell University, Ithaca, NY 14853}
\email{emv7@cornell.edu}

\date{\today}

\begin{abstract}
We present an algorithm for generating random networks with arbitrary degree distribution and Clustering (frequency of triadic closure). We use this algorithm to generate networks with exponential, power law, and poisson degree distributions with variable levels of clustering. Such networks may be used as models of social networks and as a testable null hypothesis about network structure. Finally, we explore the effects of clustering on the point of the phase transition where a giant component forms in a random network, and on the size of the giant component. Some analysis of these effects is presented. 
\end{abstract}

\maketitle

\section{Introduction}
Numerous random network models have been proposed to replicate important aspects of the topology of real-world networks~\cite{newmanstrogatzwatts,newmanwattsstrogatz,albertbarabasi,ba00,ba02,RB03,CCDM02,smallworld1,smallworld2,newmanreview,ASBS00,dorogovMendes,KR01}. In particular, much attention has been paid to the degree distribution and the clustering coefficient. A great deal of progress has been made on network models which combine certain degree distributions with some level of clustering~\cite{davidsen,girvanGrowingSN,dorogovMendes,mendes,HK02b,gu03}.  It has been an open problem to combine these two topologies in the most general way. Is it possible to have a network model which is flexible enough to accommodate any combination of degree distribution and clustering? In this article we propose such a model and demonstrate its effectiveness by generating networks over a wide range of parameters.  

Random network models have fallen in several broad categories. Some models have focused on Monte Carlo techniques to reproduce a specific topology~\cite{newmanstrogatzwatts,newmanwattsstrogatz,newmanalgorithm}. Other models have focused on plausible mechanisms for creating a network, such as preferential attachment, while some models have specific topologies built into them (e.g. regular lattices) in order to explicate the so-called "small-world" problem~\cite{smallworld1,smallworld2,davidsen,newmanreview,ASBS00}. The model proposed here lacks the intuitive appeal of mechanism-based models, but also bears the most resemblance to this category. In common with most mechanism-based models, we produce our networks by growing them from one initial node. Most network growth models have been motivated by plausible mechanisms about how nodes enter into a network and form links. We find that being able to construct a network one node at a time also offers sufficient flexibility to combine arbitrary degree distributions and clustering.


Given a network model which can combine arbitrary degree distributions and clustering, it is of great interest to explore the effects of these parameters on quantities such as the size of the giant component and the point of the phase transition where a giant component forms. This is true with regard to clustering in particular, as so far models capable of interpolating between extremes of this parameter have been lacking. In section~\ref{sec:results} we explore the effects of clustering on the size of the giant component and point of the phase transition. In section~\ref{sec:pt} we present some analysis of our observations.

Throughout this article we will rely on the following definitions:
The \emph{degree distribution} of a network describes how many neighbors a node in a network has. The probability of a node having degree $k$ in a network is described by the degree distribution $p_{k}$, where $p_{k}$ can take the form of any well defined discrete density function over the positive integers. Examples frequently employed in the literature are
	\begin{itemize}
		\item Poisson: $p_{k} = \frac{z^{k} e^{-z}}{k!}, k\geq 0$
		\item Power-law. For our experiments, we utilize power-laws with finite cuttoffs $\kappa$: $p_{k} = \frac{k^{-\gamma}e^{-k/\kappa}}{Li_{\gamma}(e^{-1/\kappa})}, k\geq 1$ where $Li_{n}(x)$ is the nth polylogarithm of x.
		\item Exponential: $p_{k} = (1-e^{-1/\lambda}) e^{-\lambda k}, k\geq 0$
		\item Empirical: The degree distribution is estimated from a sample of a network.
		\item Gaussian
	\end{itemize} 

The \emph{clustering coefficient} $C$ describes the proportion of triads in a network out of the total number of potential triads. Formally, the clustering coefficient is defined:
	\begin{displaymath}
		C={3 N_{\Delta}\over{N_{3}}}
	\end{displaymath}
where $N_{\Delta}$ is the number of triads in the network and $N_{3}$ is the number of connected triples of nodes. Note that in every triad there are three connected triples.

There is also a measure of \emph{local Clustering} given by
\begin{displaymath}
	C_{k} = \frac{N_{\Delta}(k)}{\binom{k}{2} } 
\end{displaymath}
where $N_{\Delta}(k)$ is the average number of triads connected to vertices of degree k, and $\binom{k}{2}$ is the number of potential triads connected to a vertex of degree k.  

\section{Random Network Model}
Introducing clustering into a network with a specified degree distribution is a nontrivial problem. Any method aspiring to introduce an arbitrary amount of clustering into a network must interpolate between two extremely different topologies. When clustering is 0\%, the method must reproduce pure random networks with specified degree distributions. When clustering is 100\%, there is only one configuration a network may have: each node must be connected to a small clique where every node has the same degree, and all of a node's neighbors are connected with one another. This challenge is made all the more difficult by trying to make the model networks general enough to accommodate any desired degree distribution. 

The most obvious way of introducing triads is to simply define a \emph{rewiring rule} whereby links are swapped between nodes so as to introduce triads while leaving the degree distribution the same. Such rewiring schemes quickly run into problems, as it is impossible to define a rule where the number of triads is strictly increasing and the number of triads introduced does not max out. The problem is that when links are "swapped" among nodes, triads are not only created but can be destroyed. For example, we have found that such schemes are effective only for introducing about 15\% clustering into a poisson random network.

Newman~\cite{mejn03} and Guillaume et al.~\cite{gu03} have had some success with another approach. These authors define a bipartite network of individuals and affiliations. Then they project the bipartite network onto a unipartite network of only nodes and no affiliations by connecting two nodes if they share a common affiliation. The distributions of affiliation size and the affiliation-degree distribution of the nodes is chosen in such a way as to produce a desired level of clustering. Tuning the degree distribution simultaneously has proven more challenging, however. While the bipartite projection method may actually have the potential to generate pure random networks with tunable degree distributions and clustering, so far it's efficacy has only been shown for exponential and power-law random networks, and it remains an open problem to implement it for arbitrary degree distributions. 

\begin{figure}
\includegraphics[width=\columnwidth,clip=true]{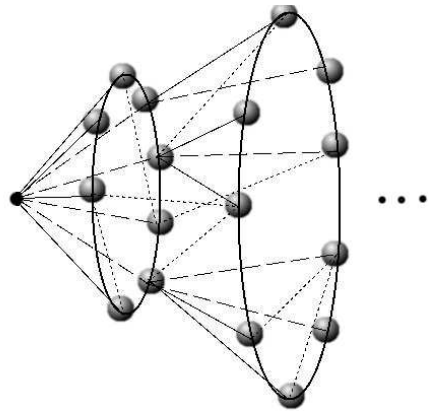}
\caption{\label{fig:schema} Overview of the network construction process. The first node (far left) is chosen at random. Then neighbors for that node are chosen as described in the text. Subsequently, neighbors are chosen for the new nodes, but now we have new connections  formed with nodes two steps away with probability C. Triadic connections are indicated with dotted lines. This process continues until the waves die out, and a new component is formed, or all nodes are exhausted.}
\end{figure}

\begin{figure*}
\begin{center}
\fbox{\includegraphics[width=.75\columnwidth,clip=true]{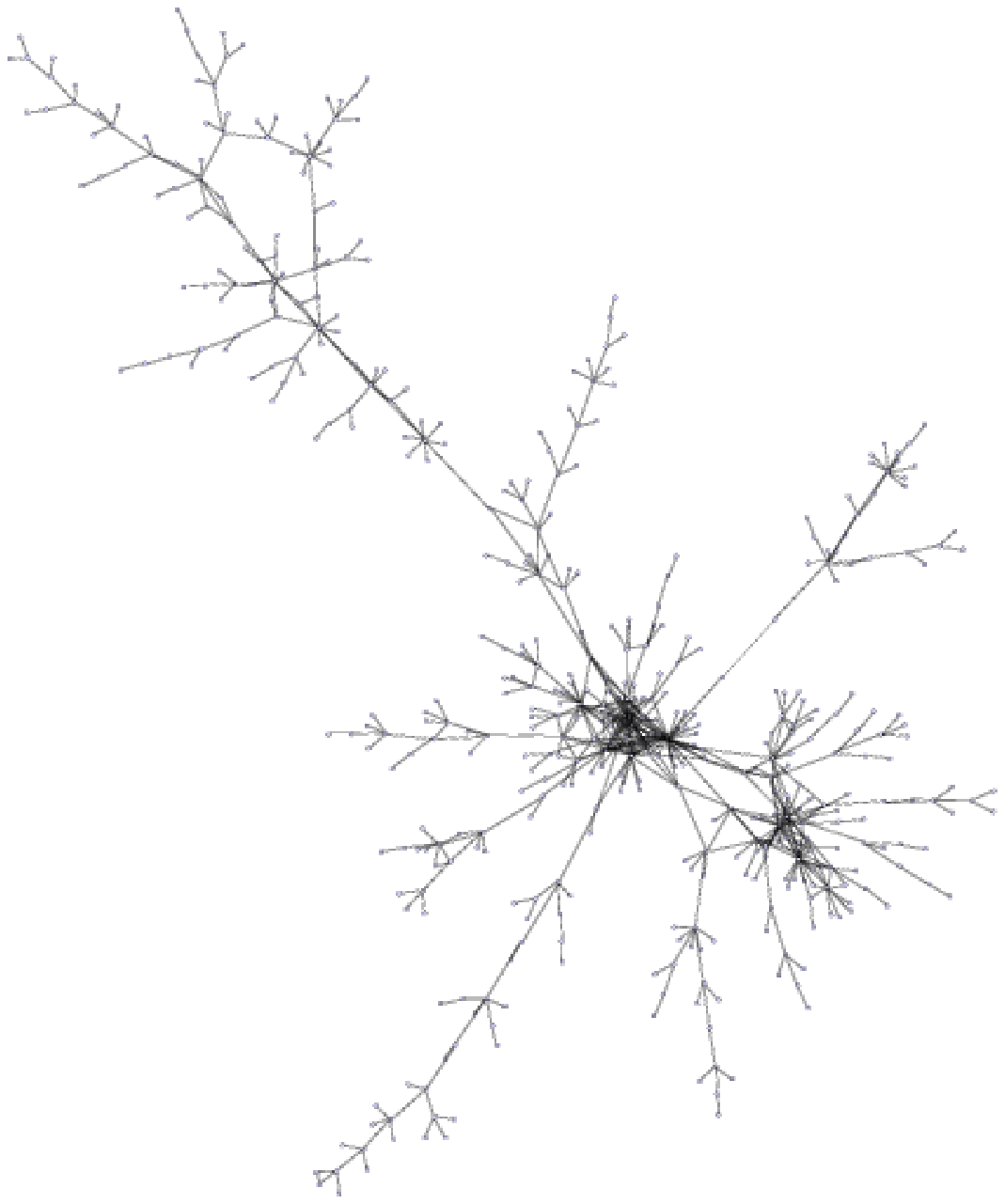}}
\fbox{\includegraphics[width=.75\columnwidth]{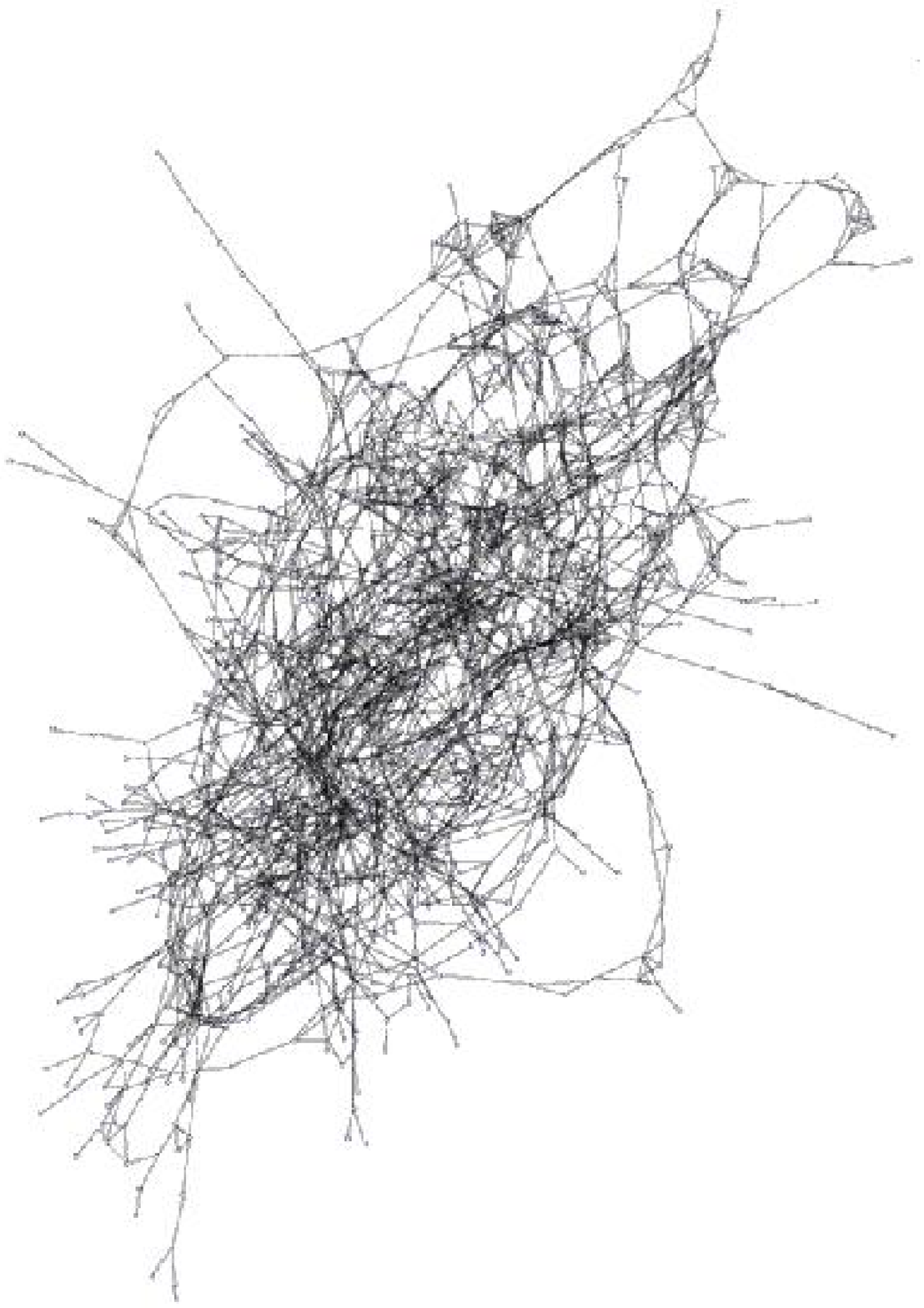}}
\caption{\label{fig:icon}Left: Random network with power law degree distribution, $\kappa = 15$, $\gamma = 2$, $C=0.15$. Right: Random network with poisson degree distribution, $z = 4$, $C = 0.40$.~\cite{yed} }
\end{center}
\end{figure*}

Our method works by growing networks. The algorithm first initializes all nodes with a degree drawn i.i.d. from the desired degree distribution. Then the random network is constructed by an iterative procedure similar to a branching process. The premise is to start from a single node and then assign new connections entirely at random under the constraint that a certain amount of clustering must exist. The algorithm is described in detail below, and is schematized in figure~\ref{fig:schema}. Two example networks are shown in figure~\ref{fig:icon}.

\begin{center}
\begin{enumerate}
\begin{small}
	\item Initialize all nodes with a degree drawn i.i.d. from the degree distribution
	\item Form a list of "stubs"-- connections of nodes which have not yet been matched  
	with neighbors. Call this list StubList.
	\item Pick a starting node, $v_{0}$, uniformly at random from all nodes. 
	\item For each of $v_{0}$'s stubs, choose a new neighbor by picking an element $v_{1}$ from the stublist
	with probability $p_{v_{1}|d(v_{0})}$ as described in the text.
	If the new neighbor is not
	\begin{itemize}
		\item the same vertex as $v_{0}$ 
		\item already connected to $v_{0}$
	\end{itemize}
	then form the connection.  
	Otherwise, repeat the process until a valid neighbor is found. 
	Add all of the neighbors gotten from this process to a list called NextWave. 
	\item Copy all elements of NextWave to a list called CurrentWave. Remove all elements  
	from NextWave. For all elements in CurrentWave: 
	\begin{enumerate}
		\item Form a list of all nodes 2 steps away; call this list PotentialTriads 
		\item For all stubs which have not been assigned neighbors 
		\begin{enumerate}
			\item Scan through PotentialTriads. With probability $C^{input}$, connect to vertex $v_{3}\in\textnormal{StubList}$.
				Remove element $v_{3}$ from the StubList. 
			\item If no neighbors were selected from PotentialTriads, select a new neighbor 
				by choosing from StubList as above.
				If the new neighbor is not in CurrentWave, and if the new neighbor is not 
				already in NextWave, add them to NextWave 
		\end{enumerate}
	\end{enumerate}
	\item Repeat the last step until NextWave is empty following an iteration. Then, if StubList 
	 is empty, the process is complete-- all connections have been formed. Otherwise,
	 start a new component by choosing
	 a new starting vertex uniformly at random from those 
	 not yet in the network.
 
\end{small}
\end{enumerate}
\end{center}

Our model has similarities and differences with other models proposed in the literature. Like the algorithm of Milo et al.~\cite{newmanalgorithm}, each node is assigned a unique degree prior to any edges being formed between nodes. But like the model networks of Barabasi~\cite{albertbarabasi}, Dorogovtsev et al.~\cite{dorogovGrowingRandNets}, and many others, the network is constructed via a growth process. The first node is chosen at random, and subsequently nodes are added to the graph by attaching them to nodes which still have stubs that have not been matched. When the new node forms its own connections, it first forms a list of all nodes which are two steps away. Then with probability $C^{input}$, that node is selected as the next neighbor. 

One complicated feature of this algorithm concerns the probability of selecting a new neighbor from the stub list. In fact, new neighbors cannot be selected uniformly at random from the stub list, as clustering implies a certain amount of degree assortativity among the nodes in the network. For example, a node connected to a degree $k$ node has $k-1$ potential triads in common with that node, and on average will have $C(k-1)$ common triads. This implies that the node must have on average a degree at least equal to $C(k-1)$.

Because triads are distributed uniformly throughout the network, the number of triads connected to a vertex of degree $k$ is distributed $binomial(\binom{k}{2}, C)$. As noted above the number of common triads with a neighbor of degree $k$ is distributed $binomial(k-1, C)$. Let $\tau_{ij}$ denote the number of triads node $i$ has in common with node $j$, and $\tau_{ji}$ denote the number of triads $j$ has in common with $i$. Of course these two random variables should be equal. We can calculate the probability of these two potential neighbors as having an equal number of common triads as: 
\begin{displaymath}
p^{c}_{ij} = \sum_{x=0}^{min\{d(i),d(j)\}} p(\tau_{ij}=x) p(\tau_{ji} =x)
\end{displaymath}
Let $q_{j}$ denote the probability of selecting node $j$ from the stub list. Then the correct probability for selecting node $j$ as a neighbor is:
\begin{displaymath}
q_{ij} = \frac{q_{j}p^{c}_{ij}}{\sum_{\alpha}p^{c}_{i\alpha}}
\end{displaymath}
which is just $q_{j}$ weighted by the probability of the two neighbors having a compatible number of triads in common.

In order to sample from this distribution, we use Markov Chain Monte Carlo techniques. For a large number of iterations we select a new node $\beta$ from the stub list, then with probability $a_{\alpha\beta}$ we accept this new neighbor, where $\alpha$ is the currently selected node in the markov process, and 
\begin{displaymath}
a_{ij} = \frac{p^{c}_{i\mu}}{p^{c}_{i\alpha}}
\end{displaymath}
If $\beta$ is not accepted, we keep $\alpha$ for the next iteration.
The final neighbor is the node selected at the last iteration. 

\begin{figure}
\begin{center}
\fbox{\includegraphics[width=.75\columnwidth]{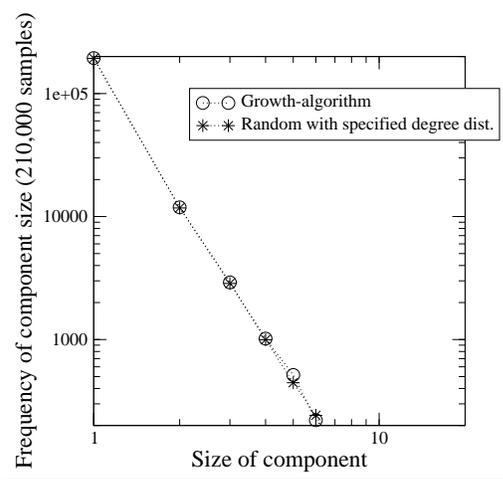}}
\caption{\label{fig:compdist}Random graphs were generated with an exponential degree distribution ($\lambda = 1.5$) with two algorithms: 1. The clustering algorithm described in this text with $C=0$ 2. A "stub-matching" algorithm as in~\cite{newmanwattsstrogatz}, known to produce true random graphs with specified degree distributions. The frequency of component sizes is illustrated above. }
\end{center}
\end{figure}

It is desirable that our algorithm produce graphs which select networks as uniformly as possible from the ensemble of all networks under the constraint of realizing a given degree distribution and clustering coefficient. It is difficult to prove that our algorithm is truly unbiased in this sense, though our networks do have many of the properties of an unbiased random network.
The algorithm produces exactly the right proportion of triads to triples in the limit of large graph size. Furthermore, the degree of the nodes were chosen as i.i.d. random variables, so in the limit of large graph size, the degree distribution is unbiased too. Furthermore, the triads are uniformly distributed throughout the network as reflected by the fact that the local clustering is independent of degree. Lastly, when this algorithm is used to produce networks with no clustering at all, it produces networks with the same statistical properties as true random graphs with a specified degree distribution. As shown in figure~\ref{fig:compdist}, the distribution of component sizes for networks made with this algorithm is identical to true random graphs with specified degree distribution without clustering.

It is worth noting that many real-world networks, particularly in the biological realm, have local clustering which scales as $1/k$~\cite{Ravasz02}. Our model in contrast produces constant local clustering, though it may be possible to generalize our method to create networks with any desired schedule of local clustering. 

\section{Results}
\label{sec:results}

\begin{figure}
\begin{center}
\fbox{\includegraphics[width=.75\columnwidth]{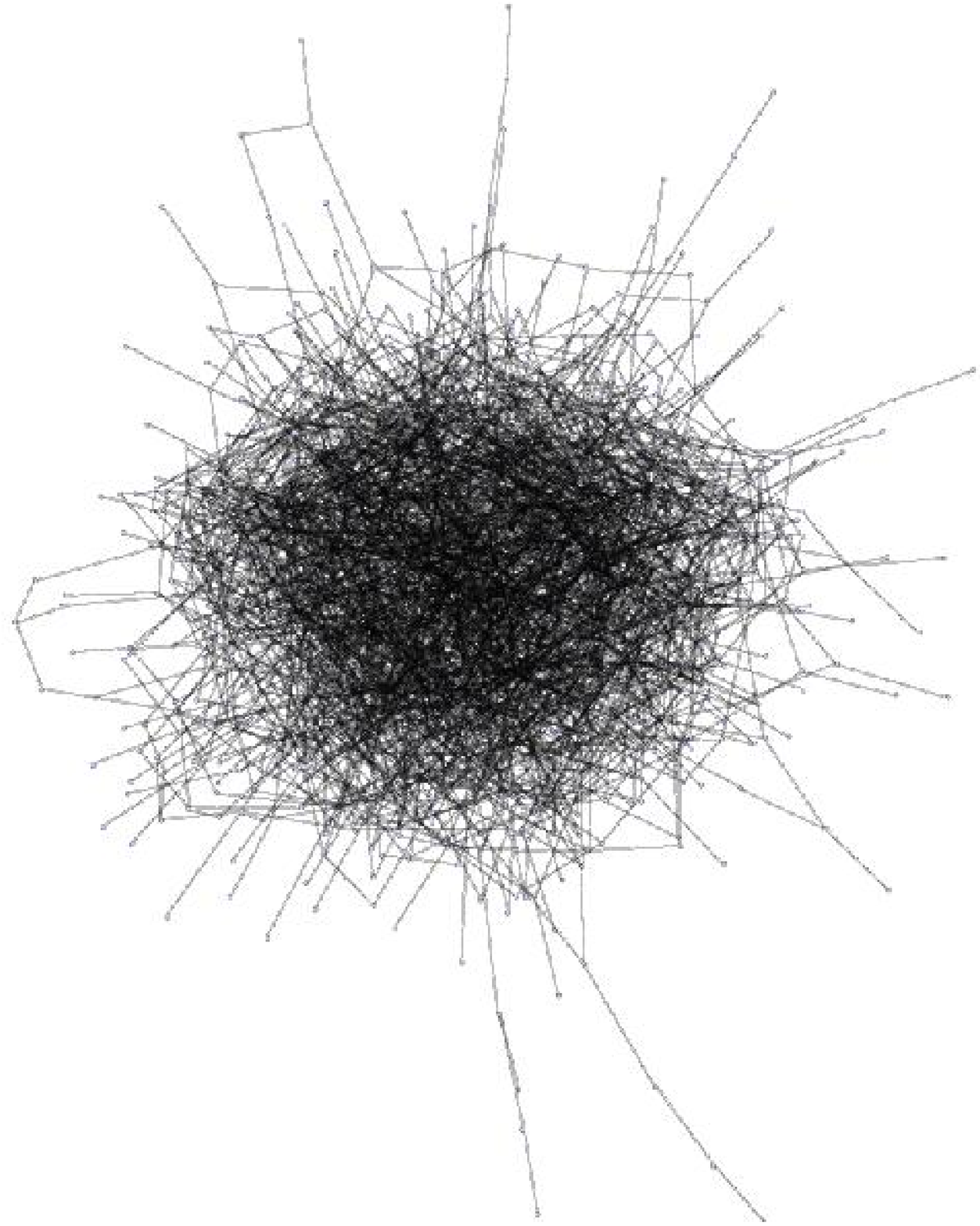}}
\caption{\label{fig:spectrum1}Random network on 1500 nodes, poisson degree distribution (z = 4), C = 0.00}
\end{center}
\end{figure}

\begin{figure}
\begin{center}
\fbox{\includegraphics[width=.75\columnwidth]{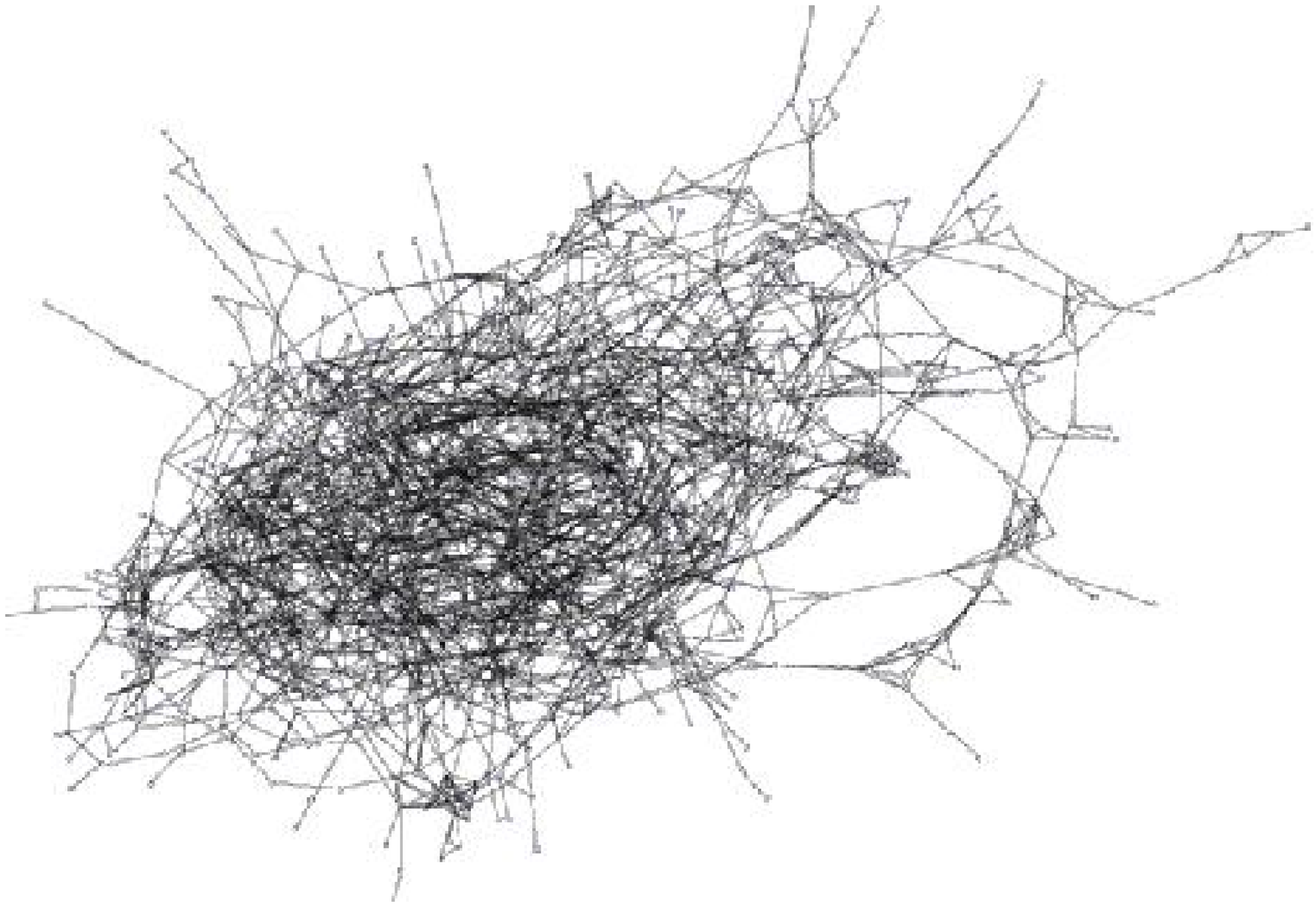}}
\caption{\label{fig:spectrum2}Random network on 1500 nodes, poisson degree distribution (z = 4), C = 0.30}
\end{center}
\end{figure}

\begin{figure}
\begin{center}
\fbox{\includegraphics[width=.75\columnwidth]{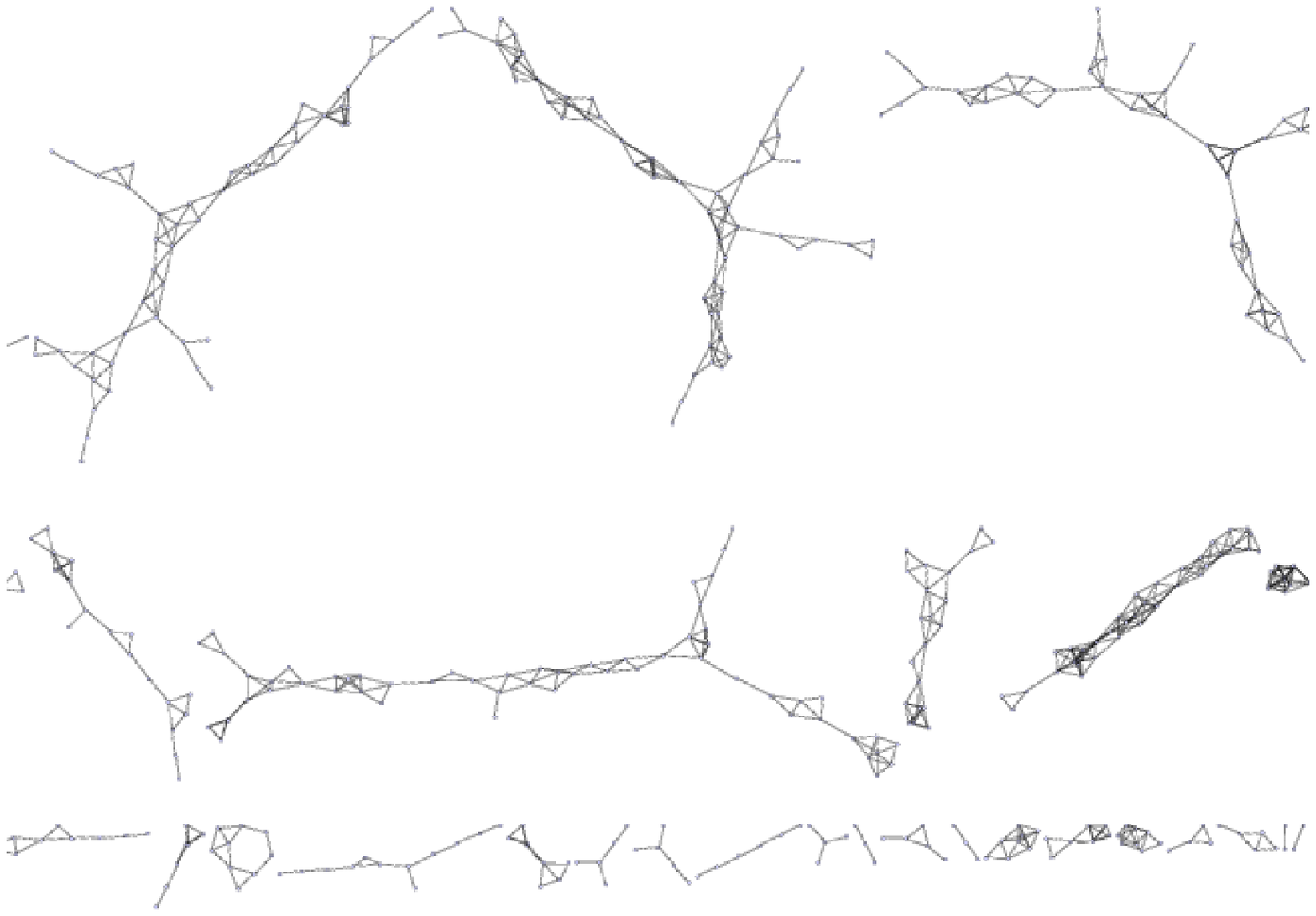}}
\caption{\label{fig:spectrum3}Random network on 1500 nodes, poisson degree distribution (z = 4), C = 0.60. The image is zoomed on several of the largest components.}
\end{center}
\end{figure}

\begin{figure}
\begin{center}
\fbox{\includegraphics[width=.75\columnwidth]{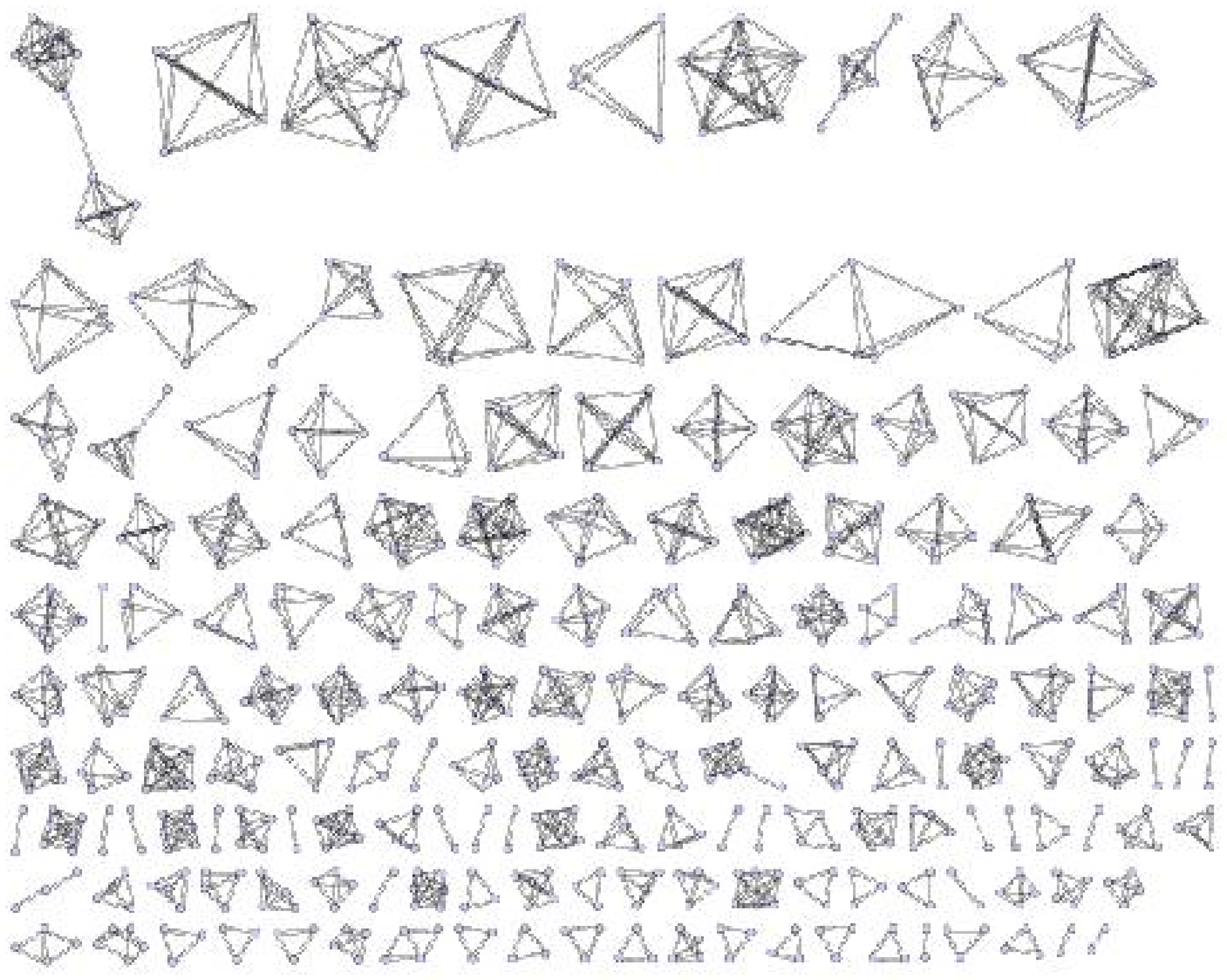}}
\caption{\label{fig:spectrum4}Random network on 1500 nodes, poisson degree distribution (z = 4), C = 0.97}
\end{center}
\end{figure}

We have explored the effects of clustering and degree distribution over a wide range of parameters. Figures~\ref{fig:spectrum1} through~\ref{fig:spectrum4} illustrate the effect of clustering on the structure of a random networks with poisson degree distributions ($z = 3$) as clustering is increased from 0 to 1.00. As clustering increases, nodes tend to disaggregate into smaller tightly connected clusters of nodes with similar degree. This has the overall effect of decreasing the giant component size as clustering is increased. In the limit as C goes to 1, we find that the network breaks down into many small completely connected cliques with each node in a clique sharing a common degree. 

\begin{figure}
\begin{center}
\includegraphics[width=.75\columnwidth,clip=true]{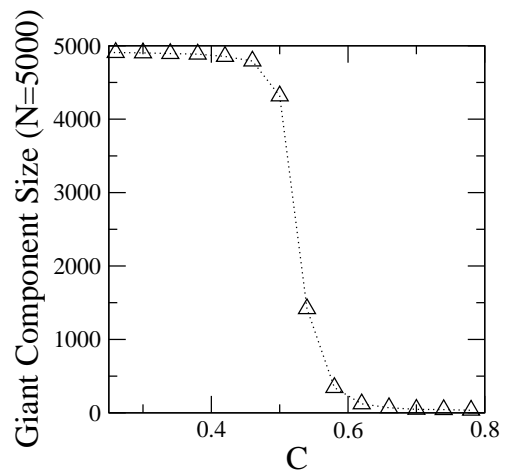}
\caption{\label{fig:ce1pois}Size of the giant component versus the clustering coefficient in a poisson random network, $z = 3$. Each point represents the average of 40 trials.}
\end{center}
\end{figure}

\begin{figure}
\begin{center}
\includegraphics[width=.75\columnwidth,clip=true]{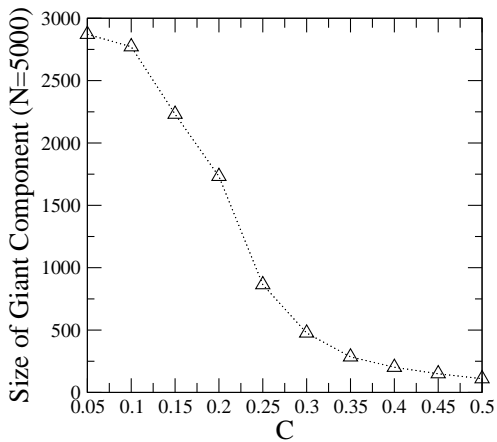}
\caption{\label{fig:ce1}N=5,000 nodes. Power law with parameters $\kappa=10$ and $\gamma$ = 2. Each point represents the average of 40 trials. Contrast this with~\ref{fig:ce1pois}. The phase transition is much less sharp than for the poisson random networks. }
\end{center}
\end{figure}

Figure~\ref{fig:ce1pois} shows the effects of clustering on the size of the giant component for a poisson random network. Clustering varies from 0.05 to 0.90. The giant component seems to undergo a phase transition at a critical level of clustering around $C=0.60$. In the next section we will find that the critical clustering value is actually $C^{*}=0.618$. At this point, nodes suddenly disaggregate into much smaller, tightly inter-connected groups. Similar phase transitions have been observed throughout the networks literature, particularly concerning the targeted deletion of links and nodes in percolation phenomena~\cite{st}. This algorithm has similar disconnecting results without modifying the degree distribution of the network.

\begin{figure}
\begin{center}
\includegraphics[width=.75\columnwidth,clip=true]{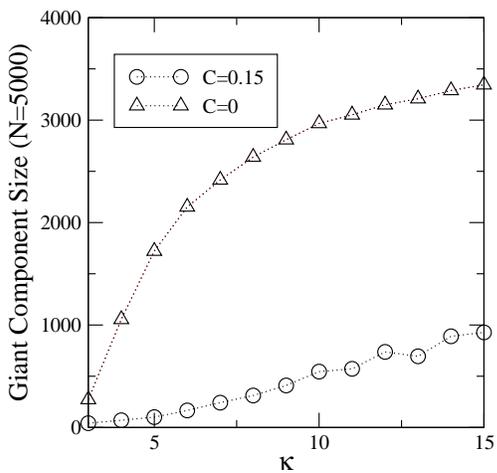}
\caption{\label{fig:ce2trial2}Two random networks are compared over a range of parameter values for the power law degree distribution with parameters $\kappa$ and $\gamma = 2$. Each point represents the average of 40 trials.}
\end{center}
\end{figure}


Regarding power-law networks (see figure~\ref{fig:ce1}), we note the striking tendency for moderate levels of clustering to inhibit the formation of the giant component. Because the number of potential triads connected to a node scales as $k^{2}$, the high degree vertices account for most of the clustering. In networks with highly skewed degree distributions such as power laws, the high-degree nodes must connect to one another in order to realize the required number of triads. This has the effect of limiting the ability to act as hubs for low-degree vertices, and consequently the network disconnects into smaller components. Large components can be preserved under much higher clustering with distributions such as the poisson. 

The phase transition also undergoes major changes with the introduction of clustering, although this effect seems to depend sensitively on the degree distribution. In figure~\ref{fig:ce2trial2} we see that the phase transition where a giant component forms is not significantly affected by the introduction of clustering for networks with power law degree distributions. In contrast to the poisson random networks, there is no sharp phase transition between the regime with a giant component and without. This bears some resemblance to percolation phenomena, where the phase transition disappears for true power-laws and an exponent of 2. But in figure~\ref{fig:e3pois} we see that the point of the phase transition was dramatically shifted forward for the poisson random network. It is somewhat surprising to observe the phase transition being shifted \textit{forwards} as our algorithm features the introduction of degree assortativity into the network. Previous research has shown the tendency of degree assortativity to shift the point of the phase transition backwards~\cite{newmanDegAss}.

\section{Phase transitions}
\label{sec:pt}
By \emph{giant component} we mean a component which in the limit of large network size occupies a proportion of the nodes greater than zero. The \emph{phase transition} is a manifold in the parameter space of $C$ and the parameters governing the degree distribution where a giant component comes into existence. It is a necessary condition for a giant component to exist that if we pick a node at random, the average number of neighbors two steps away, $s_{2}$, exceeds the number of neighbors one step away, $s_{1}$~\cite{moRe95}. This is intuitive, since if it were not the case, the number of neighbors $n$ steps away would decrease to zero on average, and the component would be finite in the limit of large network size. 

We can use this to approximate the point of the phase transition. Formally, we will solve for the point where
\begin{equation}\label{eqn:balance}
s_{1}=s_{2}
\end{equation}
The necessary condition~(\ref{eqn:balance}) will not quite be a sufficient condition in the presence of clustering as described below. Thus, our solution will only be a lower bound on the point of the phase transition, but in practice, this will serve as an excellent approximation.

For the poisson degree distribution, the average number of nodes one step away is equal to the parameter of the distribution $z$, so we have $s_{1}=z$. As is well known~\cite{newmanstrogatzwatts}, the number of edges emanating from a node if we pick an edge at random and follow it to one of its ends is also $z$ for the poisson degree distribution. Thus, in the absence of clustering we would have simply $s_{2} = s_{1}z = z^{2}$, where $s_{2}$ is the average number of nodes two steps away from a randomly chosen node. 

In the presence of clustering, things become more complicated. Lets pick a node uniformly at random in the network and call this node $v_{0}$. A neighbor of this node, $v_{1}$ will have on average $z$ connections not in common with $v_{0}$. Furthermore, there will be on average $Cz$ triadic connections between $v_{0}$ and $v_{1}$ as each of those connections has a probability $C$ of being a triad. We can simply deduct the triadic connections from $s_{2}$, so that we have
\begin{equation}\label{eqn:ub}
s_{2} > z^{2} - C z^{2} = z^{2}(1-C)
\end{equation} 
There is not equality in equation~\ref{eqn:ub} because there is an additional force limiting the number of second neighbors: Once two neighbors of $v_{0}$, say $v_{1}$ and $v_{1}'$ share a triadic connection, it becomes more likely that a node two steps away from $v_{0}$, say $v_{2}$, is a common neighbor of both $v_{1}$ and $v_{1}'$. In fact, such connections exist with probability $C$. Then, the number of connections we should deduct from every neighbor at distance two due to common connections of nodes at distance one is equal to $C$ times the average number of triadic connections at distance one, or in other words $z^{2}C^{2}$. Thus, we have
\begin{displaymath}
s_{2} = z^{2} - Cz^{2} - C^{2}z^{2} = z^{2}(1-C-C^{2})
\end{displaymath}
We can use this to solve for the critical $z^{*}_{C}$ where a giant component forms given a level of clustering $C$:
\begin{equation}
z = z^{2}(1-C-C^{2})
\end{equation}
The non-zero root of this equation is given by
\begin{equation}\label{eqn:ptsol}
z^{*}_{C} = \frac{1}{1-C-C^{2}}
\end{equation}
Note that when C=0, we retrieve the well known result that a giant component forms when $z=1$ in the absence of clustering. Unfortunately, we can only say that this is a lower bound for the phase transition due to that the nodes at distance two are not identical to $v_{0}$. The number of outgoing connections from such nodes (to nodes not already counted) is less than $z-C^{2}z$ on average. 

\begin{figure}
\begin{center}
\includegraphics[width=.9\columnwidth,clip=true]{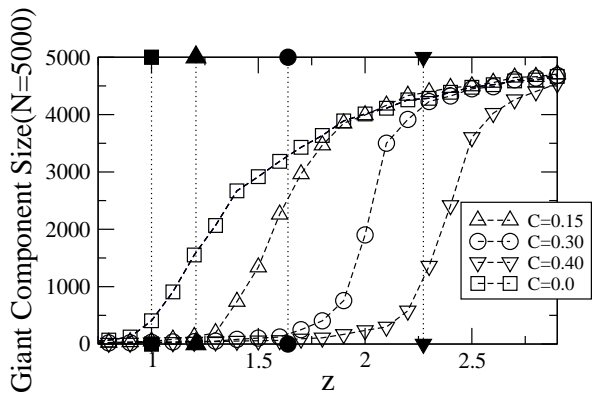}
\caption{\label{fig:e3pois}The size of the giant component is shown vs. $z$, the parameter of the poisson degree distribution, for four levels of clustering ($C=0.0, C=0.15, C=0.30, C=0.40$). The vertical lines indicate the point of the phase transition for each level of clustering predicted by equation~\ref{eqn:ptsol}}
\end{center}
\end{figure}

In figure~\ref{fig:e3pois} we have plotted the size of the giant component versus the parameter $z$ for several levels of clustering. The vertical lines correspond to the phase transitions $z^{*}_{C}$ as given by~(\ref{eqn:ptsol}). We find good agreement between theory and simulation.

There is a singularity in~(\ref{eqn:ptsol}) where $1-C-C^{2}=0$. At this point, $C^{*} = 0.618$, the giant component disappears regardless of the average degree $z$ of the degree distribution. $C^{*}$ represents the critical level of clustering that can coexist in a network with a giant component. 

\section{Finite size effects}
\label{sec:finiteSizeEffects}
During the execution of the algorithm, it occasionally happens that a node cannot find a suitable neighbor due to the absence of a node left in the network which has the correct degree and free stubs to satisfy the degree assortativity requirements. This imperfection is due to the finite size of the network. In the limit of large size, it would always be possible to find a scale such that every node can find just the right profile of neighbors with the right degree. There is no perfect way to deal with such discprepancies. For the simulations used in this article, we have simply truncated the degree of that node so that it does not have to seek a new neighbor. Even with networks of only 5000 nodes, the number of corrections made is quite small.

\begin{figure}
\begin{center}
\includegraphics[width=.75\columnwidth,clip=true]{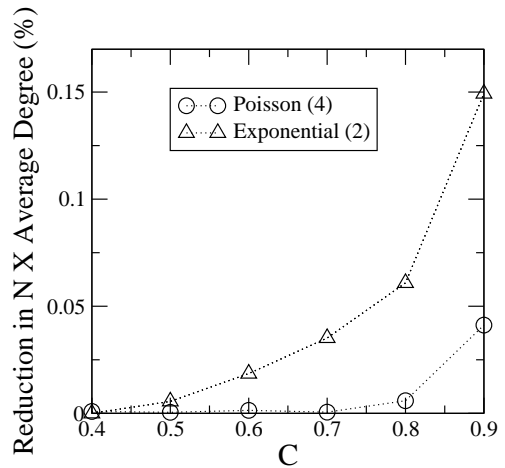}
\caption{\label{fig:trialsc1}The percentage reduction in the number of "stubs" is shown versus the Clustering Coefficient for two networks: (i) Poisson degree distribution with parameter = 4, (ii) Exponential degree distribution with parameter = 2. N=5000 for both networks. Each point is based on the average of 20 trials.}
\end{center}
\end{figure}

\begin{figure}
\begin{center}
\includegraphics[width=.75\columnwidth, angle=0]{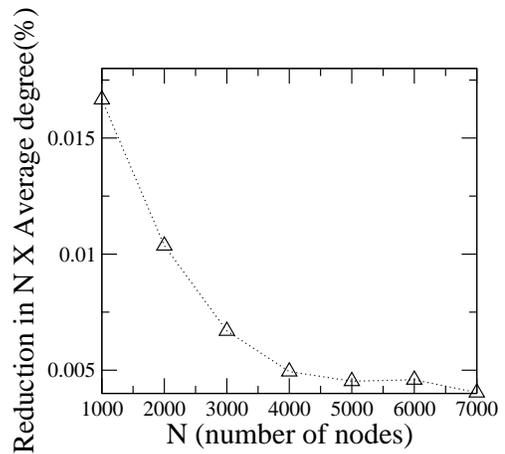}
\caption{\label{fig:trialsc2}The percentage reduction in the number of "stubs" is shown versus the network size. The network has a Poisson degree distribution with parameter = 4, C = 0.80. Each point is based on the average of 20 trial networks. }
\end{center}
\end{figure}

Figures~\ref{fig:trialsc1} and~\ref{fig:trialsc2} show the effects of network size and clustering on the amount of degree-corrections made by the algorithm. Figure~\ref{fig:trialsc1} shows the effects of clustering on the number of corrections made for two networks. Note that the total number of "stubs" in the network is equal to the average degree of the nodes times the population size. The corrections made is shown as the percent reduction in the number of "stubs". Even at 90\% clustering, the poisson random network only undergoes less than 5\% reduction in its "stubs". 

Figure~\ref{fig:trialsc2} shows the effects of network size on the number of corrections made. As expected, the number of corrections drops with the number of nodes in the network. For 7000 nodes and 80\% clustering, a poisson random network undergoes less than a 0.5\% reduction in its "stubs". 

\section{Discussion}
We have presented a method for generating random networks which unite two frequently modeled topological features-- clustering and  the degree distribution. Our model allows networks to be generated over the full spectrum of combinations of these parameters. 

Random network models can serve several important purposes. First, they can serve as a null hypothesis about the structure of a real-world network. Significant deviations in the structure of the real-world network from a corresponding random graph indicate that there are more forces at work shaping the network than are being accounted for in the random graph model. These deviations can then motivate further inquiry into the forces shaping real-world networks~\cite{newmanstrogatzwatts}.

Secondly, real-world networks are very often of a scale that it is impossible to map them entirely. Various network sampling techniques have been devised to estimate features of the network topology in the absence of data on the entire network~\cite{doug,dougsalganik,chavez}. Given reliable estimates about network topology, a random network can then be generated which reproduces this topology. The random network may be used as a stand-in for modeling various dynamic models on networks.

Lastly, the family of random networks we have presented here enables the exploration of a huge parameter space for models on networks.  There are a growing number of models which describe dynamic processes explicitly on networks. Examples are models of diffusion processes, such as models of epidemics~\cite{ancelmeyers,kretMorris,coErAvHa01}, models of fads~\cite{djwCascades,damon}, the spread of rumors~\cite{rumor1,rumor2}, and the migration of species among connected habitats~\cite{ellner}. Other models explore reciprocal interactions among nodes embedded in a network. Examples include spin-glasses, kuramoto oscillators, and disordered neural networks. There are numerous potential applications for exploring the effects of clustering and degree distributions on these and other models.

\end{document}